\documentstyle[editedvolume,epsf,numreferences]{crckapb}

\def\gtorder{\mathrel{\raise.3ex\hbox{$>$}\mkern-14mu
             \lower0.6ex\hbox{$\sim$}}}

\def\ltorder{\mathrel{\raise.3ex\hbox{$<$}\mkern-14mu
             \lower0.6ex\hbox{$\sim$}}}

\def\apj{ApJ\ }

\begin{opening}
\title{Stellar Dynamics of Dense Stellar Systems}

\author{Piet Hut}
\institute{Institute for Advanced Study,
Princeton, NJ 08540, USA}

\end{opening}

\runningtitle{Dense Stellar Systems}

\begin{document}


\begin{abstract}
Stellar dynamics is almost unreasonably well suited for an
implementation in terms of special-purpose hardware.  Unlike the case
of molecular dynamics, stellar dynamics deals exclusively with a
long-range force, gravity, which leads to a computational cost scaling
as the square of the number of stars involved.  While special tricks
can lead to a reduction of this cost from $\sim N^2$ to $\sim N\log N$
in the case of very large particle numbers, such tricks are not
suitable for all areas within stellar dynamics.  When a stellar system
is close to equilibrium, and has a very high density, it still pays to
compute all interactions on a star by star basis, even for $N=10^5$.
Any $cN\log N$ approach would either gloss over the subtle net effects
of near-canceling interactions, driving the evolution of such a
system, or would carry a prohibitively large coefficient $c$.  This
paper presents a brief introduction to the stellar dynamics of dense
stellar systems, aimed at researchers using special purpose computers
in other branches of physics.
\end{abstract}

\section{Introduction}

Stellar dynamics is the branch of astrophysics that studies the
structure and evolution of collections of stars, from small groups to
larger star clusters to entire galaxies and clusters of galaxies.  The
interactions between the individual stars can be modeled to a high
degree of accuracy as Newtonian gravitational interactions between
point masses.  Only under extremely high densities, such as occurs in
the nuclei of galaxies and the centers of the densest star clusters,
do stars have a reasonable chance to undergo a physical collision
during their life time.  In contrast, a typical star, such as our own
Sun, has a probability of only 1 in $10^{8}$ to undergo a collision
with a neighboring star, during the remaining $5\times10^9$ years of
its life.

The most spectacular example of a dense stellar system within our own
galaxy is the agglomeration of roughly a million stars within the
inner parsec from the center (a parsec is a unit of length, equal to a
few light years, and corresponds to a typical distance between stars
in the solar neighborhood).  These stars describe orbits around the
black hole that resides in the very center of our galaxy.  The black
hole itself has a mass that is a few million times larger than the
mass of the Sun.  The density of stars around the black hole is a
million times larger than the stellar density in a typical part of the
galaxy, such as where we reside.  A detailed stellar dynamical
modeling of the center of our galaxy is still difficult, partly
because the observations of this heavily obscured area have only
recently become accurate enough to tell us the physical
characteristics of the system, partly because of the interference of
other physical effects, such as the presence of gas clouds and ongoing
star formation.

Before tackling the stellar dynamics of the nucleus of our galaxy, it
is therefore prudent to start our attempts with a simpler system, such
as is provided by the core of a dense globular cluster.  While most of
the stars in and around our galaxy are spread out throughout the disk,
and to a lesser extent through the halo, there are more than one
hundred isolated star clusters circling the galaxy, each containing of
order $10^6$ stars.  In a dozen or so of those globular clusters, as
they are called because of their appearance, the central densities
rival that of the density in the nucleus of our galaxy.  However, the
absence of a large black hole, as well as gas clouds and concomitant
star formation, makes it far easier to study and model globular
cluster cores in detail.  In addition, recent observations, notably
with the Hubble Space Telescope ({\it cf.} \cite{Guh96}), have
resolved those cores into individual stars, something that has not
been possible with ground-based observations.  This paper sketches
some of the progress made in the study of globular cluster cores,
emphasizing the role of special-purpose computers.

\section{Gravitation}

The gravitational $N$-body problem, stated mathematically, is the
challenge to solve the following set of coupled nonlinear ordinary
differential equations:
$$
{d^2\over dt^2} {\bf r_i}(t) = - G \sum_{j\neq i} {m_j \over | {\bf r}_i
- {\bf r}_j |^3} ({\bf r}_i - {\bf r}_j).\eqno (1)
$$
for the set of $N$ three-dimensional vectors ${\bf r}_i(t)$ as a
function of time $t$, given the $N$ constants $m_i$ and the additional
constant $G$.  Since the absolute value of a vector ${\bf v}$ is given
in terms of its components $(v_1, v_2, v_3)$ as $|{\bf v}|=
\sqrt{v_1^2 + v_2^2 + v_3^2}$, we have here a set of $3N$ differential
equations over the real numbers, or alternatively a single
differential equation for a state vector in $3N$-dimensional real
space.

Interpreted physically, these equations describe the acceleration
exerted on each particle by the Newtonian gravitational attraction of
all other $N-1$ particles.  Each of these particles stands for a
physical body, approximated here as a mass point with mass $m_i$ and
position vector ${\bf r}_i$ with respect to the center of an
arbitrarily chosen inertial coordinate system.  The overall constant
is Newton's gravitational constant $G = 6.67 \times 10^{-8}$ cm$^3$
g$^{-1}$ sec$^{-2}$.  This is the only `coupling constant' in the
system, and it is most convenient to switch to a choice of units in
which $G=1$.  After this choice, there remain two free choices, from
among the three allowed by a rescaling of mass, length, and time
scales.  These two are in practice often used so as to provide a total
system size and system mass of order unity.

The fact that a self-gravitating system of point masses is governed by
a law with only one coupling constant has important consequences.  In
contrast to most macroscopic systems, there is no decoupling of
scales.  We do not have at our disposal separate dials that can be set
in order to study the behavior of local and global aspects separately.
As a consequence, the only real freedom we have, when modeling a
self-gravitating system of point masses, is our choice of the value of
the dimensionless number $N$, the number of particles in the system.
The value of $N$ turns out to determine a large number of seemingly
independent characteristics of the system: its granularity and thereby
its speed of internal heat transport and evolution; the size of the
central region of highest density after the system settles down in an
asymptotic state; the nature of the oscillations that may occur in
this central region; and to a surprisingly weak extent the rate of
exponential divergence of nearby trajectories in the system.

It may seem surprising that the single, simple set of equations (1)
forms a good first approximation for modeling many astrophysical
systems, such as the solar system, star clusters, whole galaxies as
well as clusters of galaxies.  The reason is that gravity, being an
attractive long-range force, dominates everything else in the
Universe.  The only other long-range force, electromagnetism, is
generally not important on very large scales, since positive and
negative charges tend to screen each other.  Short-range forces, such
as gas pressure, are usually only important on small scales, such as
in the interiors of stars.  On large scales, comparable to the size of
a galaxy, pressure is rarely important.  On intermediate scales we
have giant gas clouds in which both pressure and magnetic fields can
play a role.

This dominance of gravity on cosmic scales is a fortunate feature of
our Universe.  It implies that it is relatively simple to perform
detailed computer simulations of many astronomical systems.  A concern
with the much more complicated physics of other specializations in
astronomy, such as plasma astrophysics, radiative transfer, or nuclear
astrophysics, is often not immediately necessary.

\section{Stellar Dynamics}

Stellar dynamics can be defined as studying the consequences of eq. (1)
in astrophysical contexts.  Traditionally these equations were
discovered by studying the motions of the moon and planets, and for the
next few centuries they were applied mainly to planetary dynamics. 
Before the advent of electronic computers, most effort went into
developing analytical approximations to the nearly regular motion of the
planets.  This field, known as celestial mechanics, had an important
influence on developments both in physics and mathematics.  An example
is the study of chaos, which first arose as an annoying complexity
barring attempts to make long-time predictions in celestial mechanics. 

The solar system, however, is a relatively regular system.  All
planets move in orbits close to the ecliptic, and all revolve in the
same direction.  The orbits are well-separated, and consequently no
close encounters take place.  When we look around us on larger scales,
that of star clusters and galaxies, we do not encounter such regularities.
In our galaxy, most stars move in the galactic disk, revolving in the
same sense as the sun, but close encounters are not excluded.  In
globular star clusters, there is not even a preferred direction of
rotation, and all stars move as they please in any direction.  Does
this mean that analytic approximations are not very useful for such
systems?

The answer is: ``it depends on what you would like to know''.  In
general, the less regular a situation is, the larger the need for a
computer simulation.  For example, during the collision of two
galaxies each star in each of the galaxies is so strongly perturbed
that it becomes very difficult to predict the overall outcome with pen
and paper.  This is indeed an area of research which had to wait for
computers to even get started, in the early seventies
\cite{Too72}.  On the other hand, when we want to understand the
conditions in a relatively isolated galaxy, such as our own Milky Way,
then there is some scope for pen-and-paper work.  For example, a rich
variety of analytic as well as semi-numerical models has been
constructed for a range of problems related to the study of the
structure and evolution of regular galaxies \cite{Bin87}.  However,
even in this case we often have to switch to numerical simulations if
we want to obtain more precise results.

\section{Two Flavors of Stellar Dynamics}

The field of stellar dynamics can be divided into two subfields,
traditionally called collisional and collisionless stellar dynamics. 
The word `collision' is a bit misleading here, since it is used to
describe a close encounter between two or more stars, not a physical
collision.  After all, physical collisions are excluded in principle as
soon as we have made the approximation of point particles, as is
almost always done in stellar dynamics.

Collisional stellar dynamics is concerned about the long-term effects
of close (as well as not-so-close) stellar encounters.  The evolution
of a star cluster is governed by the slow diffusion of ``heat''
through the system from the inside towards the edge.  This heat
transport occurs through the frequent interactions of pairs of stars,
in a way similar to the heat conduction in the air in a room, which is
caused by collisions between pairs of gas molecules.  The main
difference here is that individual stars in a star cluster have mean
free paths that are much longer than the size of the system.  In other
words, little heat exchange takes place during a system crossing time
scale.

Collisionless stellar dynamics is the subfield of stellar dynamics in
which the heat flow due to pairwise interactions of stars is neglected.
For small systems this approximation is appropriate when we consider the
evolution of the star system on a time scale which does not exceed the
crossing time by a very large amount.  For a system with very many
particles, such as a galaxy, the collisionless approximation is
generally valid even on time scales comparable to the age of the Universe
(which is comparable to the age of most galaxies). 

\section{Life in a Globular Cluster Core}

     Fortunately, there are systems in nature which approach the
idealizations of collisional stellar dynamics to a remarkable degree.
These are the globular clusters, among the oldest components of our
galactic system, and going their own way in wide orbits around the
galaxy.  They are largely isolated from perturbing influences of the
galactic disk, and therefore form ideal laboratories for stellar
dynamics.

     In order to get a feel for the typical physical conditions in a
globular cluster, imagine that we would live on a planet circling a
star in the very core of a dense globular.  The density of stars there
can easily be as high as a million solar masses per cubic parsec.
This is a factor $10^6$ higher than in our own neighborhood, in the
part of the galactic disk where the Sun happens to reside.  Therefore,
we can get an impression of the night sky by bringing each star that
we normally see at night closer to us by a factor $10^2$.  Each star
would thus become brighter by a factor $10^4$, which corresponds to a
difference of ten magnitudes (in the astronomical system where a
factor ten corresponds to $\Delta m = 2.5$).

     The brightest stars would thus appear at magnitudes at or above
$m \sim -10$, comparable to that of the full moon.  They would be too
bright to look at directly, because their size would be so much
smaller than that of the moon (they would still look point-like).  It
would be easy to read books by the light of the night sky.  Although
this might be helpful to increase literacy on such a planet, to
develop astronomy there would be significantly more difficult than on
Earth, given the total absence of really dark nights: the glare of the
nearby stars would mean a nightmare for those trying to study other
objects, outside the home cluster.

\section{Dynamics of Dense Stellar Systems}

Most stars do not interact much with their environment, after they
leave the cradle of the interstellar cloud in which they were born.
Some stars are born single, and stay that way throughout their life,
although they may have acquired a planetary system during the late
stages of their formation.  However, most stars are member of a double
star or an even more complex multiple system (triples, quadruples,
etc.).  In such a system, when two stars are sufficiently close, all
kind of interesting interactions may take place including the transfer
of mass from one star to another, and even the spiral-in and eventual
merging of two or more stars.

Although binary star evolution is much more complex than single star
evolution, it can generally still be studied in isolation from its
wider environment.  After all, the typical separation between stars in
the solar neighborhood (itself typical for our galaxy) is some hundred
million times larger than the diameters of individual stars.  The
exception to this rule occurs in unusually dense stellar systems.
Examples are star clusters, both in the disk of the galaxy as well as
outside (the globular clusters), and the nuclei of galaxies.  At any
given time, such systems are still dilute enough to make physical
collisions unlikely, even during many crossing times, thus allowing
point mass dynamics to provide useful first-order approximations.
However, when viewed over a time scale of billions of years, such
collisions become unavoidable.

In recent years much progress has been made in the study of physical
collisions, both theoretically in computer simulations, and
observationally by looking for `star wrecks' as tell-tale signs of
violent encounters.  For example, observations with the Hubble Space
Telescope have shown us the presence of so-called blue stragglers,
right down in the center of the most crowded star clusters
\cite{Guh96}.  Blue stragglers are unusual types of stars that are at
least compatible with being the products of stellar collision.
Earlier, millisecond pulsars and X-ray binaries already have hinted to
us more indirectly about the sagas of their formation and subsequent
interactions \cite{Hut92}.

In order to interpret this wealth of observational information,
vigorous attempts are being made by several groups to make theoretical
models for the evolution of dense stellar systems.  The first step is
to determine the long-time behavior of a large system of point masses
(a million or so), a classical problem that is still far from solved,
and that has given rise to fascinating new insights, even over the
last ten years, including the confirmation of the presence of
gravothermal oscillations after core collapse \cite{Mak96} and the
discovery of the presence of mathematical chaos in the late stages of
its evolution \cite{Bre95}.

The second step is to integrate our understanding of the dynamics of a
system of point masses with the extra complexity introduced by the
non-point-behavior, in the form of stellar evolution, physical
collisions between stars, mass loss, etc.  This integration gives rise
to an extremely complex picture of the ecology of star clusters \cite{Por97}.

\section{Length/Time-Scale Problems}

Following the evolution of a star cluster is among the most
compute-intensive and delicate problems in stellar dynamics.  The main
challenges are to deal with the extreme discrepancy of length- and
time-scales, together with the need to resolve the very small
deviations from thermal equilibrium that drive the evolution of the
system.

Simultaneous close encounters between three or more stars have to be
modeled accurately, since they determine the exchange of energy and
angular momentum between internal and external degrees of freedom.
Especially the energy flow is important, since the generation of
energy by double stars provides the heat input needed to drive the
evolution of the whole system.  Since the sizes of the stars are a
factor $10^9$ smaller than the size of a typical star cluster, there
is an enormous range of distances over which the point particle
approximation is valid.  If neutron stars are taken into account, the
problem becomes even worse, and we have to deal with a factor of
$10^{14}$ instead, for the discrepancy in length scales.

The time scales involved are even more extreme, a close passage
between two stars taking place on a time scale of hours for normal
stars, and milliseconds for neutron stars.  In contrast, the time
scale on which star clusters evolve can be as long as the age of the
universe, of order ten billion years.  The result is that we are faced
with a discrepancy of time scales of a factor $10^{14}$ for normal
stars, and $10^{20}$ for neutron stars.

Sophisticated algorithms have been developed over the years to deal
with these problems, using individual time step schemes, local
coordinate patches, and even the introduction of mappings into four
dimensions in order to regularize the 3-D Kepler problem (through a
Hopf map to a 4-D harmonic oscillator) \cite{Aar85}.  While these
algorithms have been crucial to make the problem tractable, they are
still very time-consuming.

\section{The Challenge of Near-Equilibrium Calculations}

In the central regions of a star cluster, the two-body relaxation time
scale, which determines the rate at which energy can be conducted
through the system, can be far shorter than the time scale of
evolution for the system as a whole, by several orders of magnitude.
For example, in globular clusters, density contrasts between center
and the half-mass radius can easily be as large as $10^4$, which
implies a similar discrepancy in relaxation time scales.

As a consequence, thermal equilibrium is maintained to a very high
degree.  Since it is precisely the deviation from thermal equilibrium
that drives the evolution of the system, it is extremely difficult to
cut corners in the calculation of close encounters.  If any systematic
type of error would slip in here, even at the level of, say,
$10^{-6}$, the result could easily invalidate the whole calculation.
Therefore, the use of tree codes, or other $N\log N$ methods, does not
offer much of a speed-up here, since we have to keep their tuning
parameters down to a level where the $N\log N$ calculations
effectively return to an almost full $N^2$ modeling.

This handicap for simulations of dense stellar systems shows up in the
typical particle numbers used in this field, currently in the range
$N=10^4$ to $4\times10^4$, in contrast to, say, cosmological
simulations.  In the latter collisionless systems, particle numbers of
$10^6 \sim 10^7$ are not unheard of.  A second reason for this large
discrepancy in particle number is that a typical particle in a
cosmological simulation need to perform only a thousand time steps,
while traveling from its initial to its final position.  Against the
back drop of an expanding universe, such a trip covers only a small
fraction of the total size of the modeled chunk of the universe.
Stars in a globular cluster, on the other hand, move through the
system under consideration thousands of times, with a total number of
time steps that can easily exceed $10^6$ for a single particle, during
a single simulation, especially in the central regions of a cluster.

\section{Computational Requirements}

Currently, with routine type calculations, it is only feasible to
model the evolution of a globular cluster containing roughly
$N=5\times10^3$ stars, since this requires some $10^{15}$ floating
point calculations, equivalent to 10 Gflops-day, or several months
or more on a typical workstation.  
The cpu cost of a direct $N$-body calculation scales $\propto N^3$,
where the inter-particle forces contribute two powers in $N$ and the
increased time scale for heat conduction contributes the third factor
of $N$.  Therefore, a calculation with half a million stars,
resembling a typical globular star cluster, will require $\sim 10$
Pflops-day.

In contrast, the memory requirements will remain very modest.
All that is needed is to keep track of $N=5\times10^5$ particles, each
with a mass, position, velocity, and a few higher derivatives for
higher-order integration algorithms.  Adding a few extra diagnostics
per particle still will keep the total number of words per particle to
about 25 or so.  With 200 bytes per particle, the total memory
requirement will be a mere 100 Mbytes.

Output requirements will not be severe either.  A snapshot of the
positions and velocities of all particles will only take 10 Mbytes.
With, say, $10^5$ snapshot outputs for a run, the total run worth 10
Pflops-day will result in an output of only 1 Tbyte.

\section{Special-purpose Hardware}

A significant step toward the modeling of globular star clusters has
been made in 1995 with the completion of a special-purpose piece of
hardware, the GRAPE-4, by a group of astrophysicists at Tokyo
University \cite{Mak97}.  GRAPE, short for GRAvity PipE, is
the name of a family of pipeline processors that contain chips
specially designed to calculate the Newtonian gravitational force
between particles.  A GRAPE processor operates in cooperation with a
general-purpose host computer, typically a normal workstation.  Just
as a floating point accelerator can improve the floating point speed
of a personal computer, without any need to modify the software on
that computer, so the GRAPE chips act as a form of Newtonian
accelerator.

The force integration and particle pushing are all done on the host
computer, and only the inter-particle force calculations are done on
the GRAPE.  Since the latter require a computer power that scales with
$N^2$, while the former only require power  $\propto N$, load
balance can always be achieved by choosing $N$ values large enough.

For example, the complete GRAPE-4 configuration, with a speed of more
than 1 Tflops, can be efficiently driven by a workstation of 100
Mflops.  Although such a workstation operates at a speed that is lower
than that of the GRAPE by a factor of $10^4$, load balance can be
achieved for particle numbers of order $N \sim 5\times10^5$.  In
practice, calculations are not expected to exceed a particle number of
order $10^5$, since even those simulations cannot be completed in less
than a few months (note that the extreme parallelism of the GRAPE does
not allow the most efficient scalar algorithm to be implemented).

In addition, this computer can be and has been used for simulations in
a wide range of other fields of science, such as plasma physics,
molecular dynamics, the study of turbulence, and even protein folding.
The latest GRAPE project is described in this volume by Jun Makino
\cite{Mak98}.

\newcommand{\etalchar}[1]{$^{#1}$}
\newcommand{\noopsort}[1]{} \newcommand{\printfirst}[2]{#1}
  \newcommand{\singleletter}[1]{#1} \newcommand{\switchargs}[2]{#2#1}

\end{document}